\documentclass[twocolumn]{aastex631}
\usepackage{amsmath,epsfig,url}
\usepackage{graphicx,epstopdf,float,color,array}
\usepackage{rotating,multirow}

\shorttitle{AGN SOM}
\shortauthors{Sanjaripour et al.}
\usepackage{xcolor}

\begin{document}
\title{\textbf{Selection of Dwarf Galaxies Hosting AGNs: A Measure of Bias and Contamination using Unsupervised Machine Learning Techniques}}
\author[0009-0009-3048-9090]{Sogol Sanjaripour}
\affiliation{Department of Physics and Astronomy, University of California Riverside, Riverside, CA 92521, USA}
\author[0009-0009-3048-9090]{Archana Aravindan}
\affiliation{Department of Physics and Astronomy, University of California Riverside, Riverside, CA 92521, USA}
\author[0000-0003-4693-6157]{Gabriela Canalizo}
\affiliation{Department of Physics and Astronomy, University of California Riverside, Riverside, CA 92521, USA}
\author[0000-0003-2226-5395]{Shoubaneh Hemmati}
\affiliation{IPAC, California Institute of Technology, Pasadena, CA 91125, USA}
\author[0000-0001-5846-4404]{Bahram Mobasher}
\affiliation{Department of Physics and Astronomy, University of California Riverside, Riverside, CA 92521, USA}
\author[0000-0002-2583-5894]{Alison L. Coil}
\affiliation{Center for Astrophysics and Space Sciences, Department of Physics, University of California, San Diego, CA 92093-0424, USA}
\author[0000-0001-6386-7371]{Barry C. Barish}
\affiliation{LIGO Laboratory, California Institute of Technology, Pasadena, CA 91125, USA}
\affiliation{Department of Physics and Astronomy, University of California Riverside, Riverside, CA 92521, USA}

\email{sogol.sanjaripour@email.ucr.edu}

\journalinfo{\textcopyright\ 2024. All rights reserved. To be submitted to the Astrophysical Journal.}

\begin{abstract}

Identifying active galactic nuclei (AGNs) in dwarf galaxies is critical for understanding black hole formation but remains observationally challenging due to their low luminosities, low metallicities, and star formation–driven emission that can obscure AGN signatures. Machine learning (ML) techniques, particularly unsupervised methods, offer new ways to address these challenges by uncovering patterns in complex, multiwavelength data. In this study, we apply Self-Organizing Maps (SOMs) to explore the spectral energy distribution (SED) manifold of dwarf galaxies and evaluate AGN selection biases across various diagnostics. We train a $51 \times 51$ SOM on 30,344 dwarf galaxies ($z < 0.055$, $M_* < 10^{9.5} M_{\odot}$) from the NSA catalog using nine-band photometry spanning near-UV to mid-infrared. A set of 438 previously identified dwarf AGNs, selected via mid-IR color, optical emission lines, X-ray, optical variability, and broad-line features, was then mapped onto the SOM.
AGNs identified by different methods occupy distinct and partially overlapping regions in SED space, reflecting biases related to host galaxy properties. BPT-selected AGNs cluster in higher-mass regions, while AGNs selected via [O\,\textsc{i}], [S\,\textsc{ii}], He\,\textsc{ii}, X-ray, and variability diagnostics show broader distributions. WISE-selected AGNs are strongly concentrated in lower-mass regions of the SOM, consistent with previous studies. WISE-selected AGN form two distinct clumps on the SOM: one associated with bluer starburst-like systems and the other with somewhat redder galaxies exhibiting spectral features more typical of AGN activity. This separation may help disentangle true AGN hosts from starburst contaminants in WISE-selected samples.
Furthermore, AGNs selected via traditional emission-line diagnostics (e.g., BPT, [O\,\textsc{i}], [S\,\textsc{ii}], He\,\textsc{ii}), as well as broad-line and WISE methods, tend to avoid regions of the SOM associated with strong star formation activity, consistent with known selection biases. In contrast, a subset of AGNs in lower-mass galaxies occupy SOM regions indicative of high AGN luminosity relative to their stellar content, highlighting the presence of luminous AGNs in faint hosts. Our results highlight the utility of manifold learning in refining AGN selection in the low-mass regime.

\end{abstract}


\section{Introduction}

Supermassive black holes (SMBHs), expected to reside in the cores of nearly all massive galaxies, play a central role in theories of galaxy formation and evolution. Over the past decade, the active galactic nucleus (AGN) community has made substantial progress in establishing scaling relations between SMBHs and their host galaxies, uncovering feedback processes, and detecting massive accreting black holes at high redshifts. Recent discoveries, such as the population of compact, high-redshift AGNs revealed by the \textit{James Webb Space Telescope} (JWST), the so-called "Little Red Dots" \citep{Kocevski2023, Labbe2023, Harikane2023, Barro2023, Maiolino2024, Matthee2024, Kocevski2024, Greene2024,  Labbe2025}, highlight the diversity of AGN phenomena across cosmic time. Despite this progress, our understanding of how SMBHs initially formed remains limited.

The presence of SMBHs at the centers of low-mass dwarf galaxies can significantly influence feedback processes and galaxy evolution, particularly due to the shallow gravitational potential wells of these systems. This is especially important given the high space density of dwarf galaxies at high redshift, as predicted by $\Lambda$CDM models, and their subsequent evolution into more massive galaxies in the local universe \citep{Behroozi2013}. Therefore, a crucial step in resolving the origin of SMBHs is determining their demographics in the lowest-mass galaxies in the local universe.
Different theoretical models for black hole seeding make distinct predictions for the occupation fraction and mass distribution of black holes in dwarf galaxies \citep[e.g.,][]{Greene2020}, making this regime particularly important for observational constraints. However, dwarf galaxies are particularly effective at concealing their central black holes. In this low-mass regime, the gravitational sphere of influence of a black hole is typically too small to resolve with current instruments, rendering dynamical mass measurements unfeasible. Consequently, these black holes can only be identified when actively accreting. Detecting AGNs in dwarf galaxies presents significant challenges. These systems typically have elevated levels of star formation, low metallicites, and host AGNs with low luminosities and small black hole masses. Such traits reduce the effectiveness of standard AGN detection methods, including optical emission line diagnostics, X-ray emission, radio observations, and mid-infrared color selection, as discussed by \citet[and references therein]{Reefe2023}.
Consequently, observations of AGNs in dwarf galaxies are likely biased towards higher mass BHs in bulge-dominated, metal-rich systems \citep[e.g.][]{Reines2013}. Furthermore, each diagnostic method recovers a distinct subset of AGNs, often with limited overlap, making it difficult to derive complete AGN occupation fractions \citep{WasleskeBaldassare2024}.
This fragmentation in AGN selection is further emphasized by \citet{WasleskeBaldassare2024}, who demonstrate that commonly used diagnostics—such as emission line ratios, X-ray luminosity, radio excess, and mid-infrared colors, tend to identify largely non-overlapping subsets of AGNs in dwarf galaxies. Their analysis highlights the lack of a unified AGN population across different methods, underscoring the complexity of uncovering low-mass black holes in this regime. These findings imply that current AGN census efforts in dwarf galaxies are not only incomplete, but also shaped by method-dependent selection biases, reinforcing the need for multiwavelength, complementary approaches to achieve a more representative understanding of black hole occupation in the low-mass regime.

In this paper, we conduct a project to investigate the biases that affect each AGN selection method in the dwarf galaxy regime. Using machine learning (ML) tools, we aim to analyze large populations of dwarf galaxies to assess how each diagnostic technique performs across different galaxy properties. Ultimately, our goal is to calibrate AGN occupation fractions for specific populations of dwarf galaxies, enabling more accurate studies of black hole demographics in the low-mass regime.

To address this, we employ a machine learning technique known as self-organizing maps (SOMs; \citealt{Kohonen1982}) to project high-dimensional structures onto a two-dimensional grid while preserving the topological relationships within the data. This approach enables effective visualization through dimensionality reduction.
The rapid increase in data volume from current and upcoming multiwavelength surveys has made ML and deep learning (DL) indispensable for astronomical research. These tools are capable of learning complex patterns and non-linear relationships in large, heterogeneous datasets, allowing astronomers to efficiently classify sources, estimate physical parameters, and uncover rare populations. SOMs are a form of unsupervised learning that project complex datasets onto lower-dimensional grids while preserving the topology of the input space. These methods have gained popularity in astronomy for source classification, redshift estimation, and data visualization \citep{Geach2012, Masters2015, Hemmati2019b, Davidzon2019, Davidzon2022, Busch2022, Chartab2023, Torre2024}. For instance, \citet{Sanjaripour2024} applied SOMs trained on photometric data from the Cosmic Assembly Near-infrared Deep Extragalactic Legacy Survey \citep[CANDELS;][]{Grogin2011, Koekemoer2011} to examine selection biases in the MOSFIRE Deep Evolution Field (MOSDEF) survey \citep{Kriek2015}, a near-infrared spectroscopic campaign targeting galaxies at $1.4 < z < 3.8$. Their analysis revealed that AGNs in the MOSDEF sample preferentially occupy high-mass regions of the SED manifold, highlighting biases introduced by spectroscopic targeting strategies.

In this study, we apply SOMs to the NASA-Sloan Atlas (NSA; \citealt{Blanton2011}) sample of low-redshift dwarf galaxies ($z < 0.055$, $M_\ast < 10^{9.5}~M_\odot$). We train a SOM using photometric measurements from 30,344 dwarf galaxies and map 438 previously identified AGNs from the literature and compiled by  \citet{WasleskeBaldassare2024}, selected through multiple diagnostics including mid-infrared (\textit{WISE}; \citealt{Stern2012, Satyapal2014, Hainline2016}), optical line ratios (BPT; \citealt{Baldwin1981, Kewley2006, Reines2013}), broad-line detections \citep{Trump2015, Birchall2022}, and optical variability \citep{Baldassare2018, Burke2021}. This allows us to assess the distribution of AGNs in the photometric SED space and evaluate potential selection biases and overlaps.

Beyond refining AGN selection in dwarf galaxies, our approach provides a framework for systematically comparing multiwavelength AGN diagnostics, improving classification techniques, and understanding selection effects and biases across different AGN populations. Furthermore, this methodology can be extended to upcoming deep surveys, such as those from the Rubin Observatory, \textit{JWST}, and the \textit{Euclid Space Telescope}, to enhance AGN identification in low-mass galaxies at higher redshifts.

The paper is structured as follows: Section \S 2 describes the datasets used for training the machine learning model and testing its performance. Section \S 3 outlines the methodology, including the training of the SOM and the mapping of AGN candidates. Section \S 4 presents the results, comparing the distribution of AGNs in the SOM with selections from traditional non-ML methods and evaluating potential biases, and Section \S 5 provides a summary of our findings. Throughout this work, all magnitudes are given in the AB system \citep{Oke1983}.


\section {Sample Selection and Data}

In this section, we provide an overview of the datasets used for training the ML model and testing its performance. The training set consists of photometric data from the NSA catalog, while AGN candidates selected through various methods serve as test cases.

\subsection{NSA Catalog}

The NSA catalog is a compilation of low redshift galaxies ($z < 0.055$) derived primarily from the Sloan Digital Sky Survey  (SDSS; \citealt{York2000}) and supplemented with data from other sources. The catalog provides photometric and spectroscopic measurements across multiple bands, including ultraviolet from the Galaxy Evolution Explorer (GALEX), optical from SDSS, and infrared from the Two Micron All Sky Survey (2MASS) and the Wide-field Infrared Survey Explorer (WISE) \citep{Blanton2011}. The NSA catalog was constructed to improve the characterization of nearby galaxies, particularly in terms of stellar masses, structural parameters, and emission line properties, with refined photometry optimized for extended sources.  

The SDSS, which forms the backbone of the NSA catalog, is a large-scale imaging and spectroscopic survey conducted using the 2.5-meter telescope at Apache Point Observatory \citep{Gunn2006}. The survey provides five-band optical imaging ($u_{\lambda 367 nm}$, $g_{\lambda 483 nm}$, $r_{\lambda 622 nm}$, $i_{\lambda 755 nm}$, $z_{\lambda 869 nm}$) and fiber-fed spectroscopy covering 3800–9200 Å with a resolution of $R \sim 2000$. The NSA incorporates reprocessed SDSS photometry using optimized sky subtraction and deblending techniques to better capture the flux of extended galaxies \citep{Blanton2011}. Additionally, redshift measurements and spectral line diagnostics from SDSS spectra enable detailed studies of galaxy properties, including emission line classification and AGN selection.  

Following \citet{Reines2013}, we select all galaxies from the NSA catalog version \texttt{nsa\_v1\_0\_2} with stellar masses M$_*$ $<$ 3 $\times$ 10$^9$ M$_{\odot}$, which is the approximate mass of the Large Magellanic Cloud (\citealt{Marel2002}). This results in a sample of 44,494 dwarf galaxies.

\subsection{WISE Photometry}

The \textit{WISE} is a space-based infrared survey mission that has mapped the entire sky in four mid-infrared bands: W1 (3.4\,\micron), W2 (4.6\,\micron), W3 (12\,\micron), and W4 (22\,\micron) \citep{Wright2010}. Designed to detect warm dust, star formation, and AGN activity, \textit{WISE} has played a key role in identifying galaxies across a wide range of stellar masses, including dwarf systems (e.g., \citealt{Jarrett2011, Satyapal2014}). However, its sensitivity limits and resolution constraints may introduce selection effects, particularly for low-mass, low-luminosity galaxies.  

To incorporate mid-infrared data into our analysis, we cross-matched the dwarf galaxies from the NSA catalog with the \textit{WISE} All-Sky Source catalog, to obtain \textit{WISE} photometry for the dwarf galaxies. The cross-matching was done using the closest matching source to the galaxy position from the literature within a 2 $\arcsec$ 2-D sky distance, resulting in a final training sample of  30,344 dwarf galaxies with UV, optical, and mid-infrared fluxes. This cross-matching allows us to explore infrared color distributions and their connection to AGN selection in dwarf galaxies.

\subsection{AGN Selection in Dwarf Galaxies} 

Identifying AGNs in dwarf galaxies requires a multiwavelength approach, as no single method can fully capture the AGN population. In this study, we used a set of previously identified AGNs as test cases to analyze their distribution within the NSA catalog. To do so, we matched dwarf galaxies in the NSA catalog with WISE photometry to the sample of known dwarf galaxies with AGN signatures compiled by \citet{WasleskeBaldassare2024}. Their catalog includes 733 active dwarf galaxies identified using a range of diagnostic techniques, including optical spectroscopy, X-ray emission, mid-infrared colors, and optical photometric variability \citep[see][for details on individual selection methods]{WasleskeBaldassare2024}.

The two catalogs were cross-matched using an automated two-step algorithm designed to uniquely associate each entry from the \citet{WasleskeBaldassare2024} catalog with a counterpart in our compiled NSA catalog of dwarf galaxies. In the first step, to account for the limited coordinate precision in the \citet{WasleskeBaldassare2024} catalog (reported to five significant figures), the coordinates in the NSA catalog were rounded to the same precision.
Candidate matches in the \citet{WasleskeBaldassare2024} catalog were identified by comparing these rounded values, considering only entries that had not already been matched. If multiple candidates were found, the distance between the original coordinate pairs was computed, and the candidate with the smallest distance was selected as the match. For NSA catalog entries that did not yield an exact match, an iterative matching approach was applied. Starting with a tolerance of 
10$^{-6}$ degrees, candidate matches were searched by selecting \citet{WasleskeBaldassare2024} catalog entries within this tolerance in both RA and DEC. If no match was found, the tolerance was doubled, and the search repeated until a maximum tolerance of 
10$^{-2}$ degrees was reached. Again, the candidate with the smallest distance between the coordinate pairs was chosen if multiple possibilities existed. The matches with higher tolerance were also visually inspected to ensure that they were correctly matched. By employing this method, we found 438 dwarf galaxies with AGN signatures in our sample. This number is lower than the number of identified AGN in the \citet{WasleskeBaldassare2024} sample because we were limited to only the NSA catalog (and thus were limited to galaxies with z $<$ 0.055), while the AGN in the \citet{WasleskeBaldassare2024} were identified from other surveys such as the Galaxy and Mass Assembly survey (GAMA). 

By cross-matching these 438 identified AGNs and mapping them onto the trained SOM, we examine their distribution in the SED space and assess potential selection biases across different AGN identification techniques in dwarf galaxies as described below.

\begin{figure*}[htbp]
\centering
  \includegraphics[trim=0cm 0cm 0cm 0cm, clip,width=1 \textwidth] {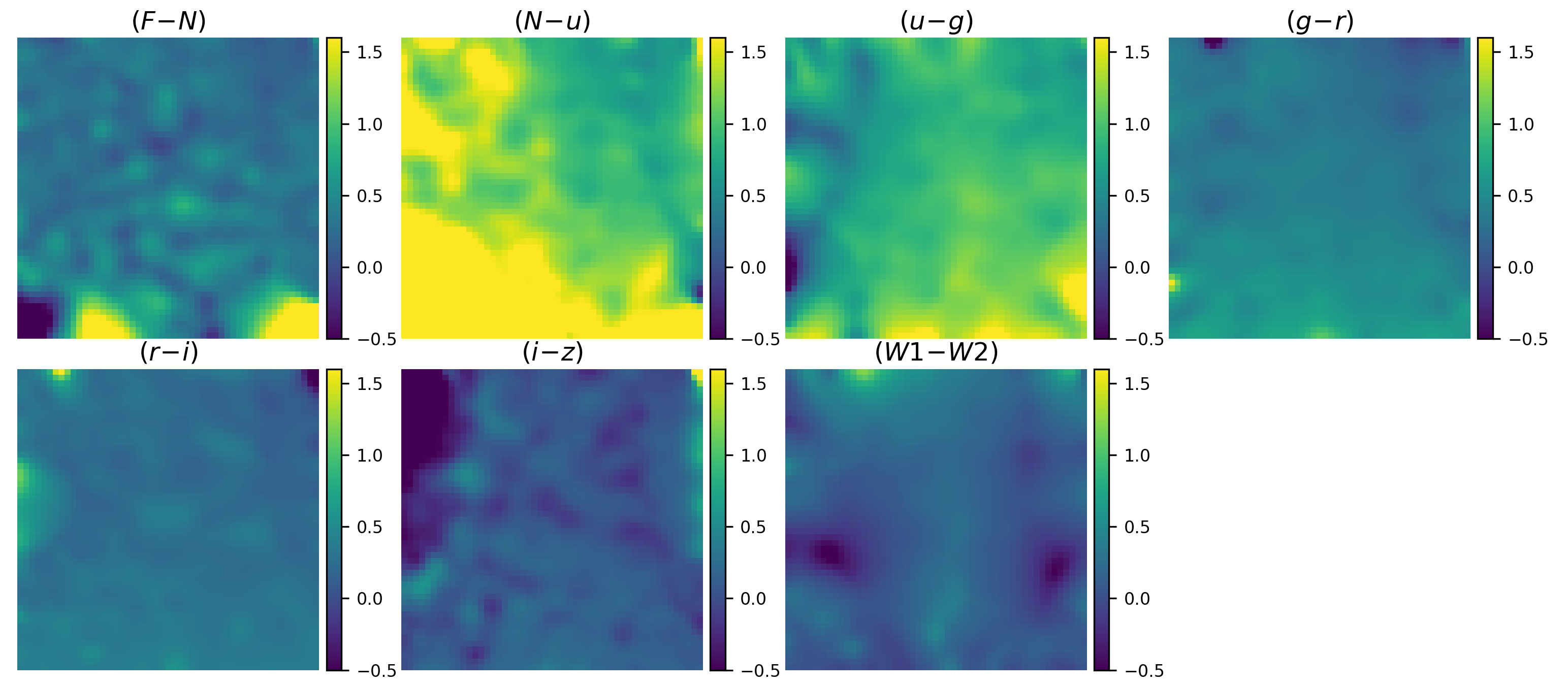}
\caption{Component maps of the trained SOM, showing the learned 7-dimensional color manifold. Each panel corresponds to one of the color dimensions, with the color bars indicating the variation in that color across the SOM cells.}

\label{fig:som_train}
\end{figure*}

\section{Method}

\subsection{Self Organizing Maps}

The SOM is a type of unsupervised ML model designed for dimensionality reduction and clustering \citep{Kohonen1982}. They work by mapping high-dimensional data onto a lower-dimensional, typically two-dimensional, grid while preserving the topological structure of the input space. Unlike traditional clustering algorithms, such as k-means, which assign data points to discrete clusters, SOM create a continuous representation, making them well-suited for exploring complex astronomical datasets.

The SOM algorithm begins by initializing a set of neurons, each associated with a weight vector of the same dimension as the input data. During training, an input data point is compared to all neurons, and the best-matching unit (BMU), the neuron whose weight vector is closest to the input, is identified based on a distance metric, typically Euclidean distance. The BMU and its neighboring neurons are then updated using a learning rate and a neighborhood function, which gradually shrink over time, allowing the map to organize itself into a structured representation of the data. Through iterative updates, the SOM groups similar inputs close together, preserving the underlying correlations and distributions in the dataset.

\subsection{Manifold learning of Galaxy SED}

To map the high-dimensional SED space of dwarf galaxies, we trained a 51 × 51 square SOM, chosen to balance resolution and computational efficiency while capturing dataset variations. The input feature space consists of seven color dimensions (shown in Fig.~\ref{fig:som_train} ), derived from nine-band photometry: FUV (153 nm), NUV (227 nm), u (355 nm), g (468 nm), r (616 nm), i (748 nm), z (893 nm), W1 (3.4 $\mu$m), and W2 (4.6 $\mu$m). Our training sample includes 30,344 dwarf galaxies from the NSA catalog with $z < 0.055$ and $M_* < 10^{9.5} M_{\odot}$.  

We implemented the SOM using the \textit{SOMPY} package \citep{sompy}, which iteratively adjusts neuron weights to capture the underlying structure of the input data. To evaluate its performance, we used the quantization error, the average Euclidean distance between each input vector and its BMU in normalized color space. The 51~$\times$~51 grid size was selected based on the elbow point in the quantization error curve, offering an optimal balance between model complexity and representation accuracy. The resulting quantization error of 0.65 is comparable to the typical photometric uncertainties in the NSA catalog ($\sim$0.1~mag per band; \citealt{Blanton2011}), indicating that the SOM reliably captures intrinsic patterns in galaxy colors without being dominated by noise. This enables robust clustering and analysis of SEDs in low-mass galaxies.

Figure~\ref{fig:som_train} presents the trained SOM grid. The SOM is color-coded to represent the final values of the seven input features used during training, reflecting the structure learned from the photometric color space. 

\begin{figure*}
\centering
  \includegraphics[trim=0cm 0cm 0cm 0cm,clip,width=0.98\textwidth] {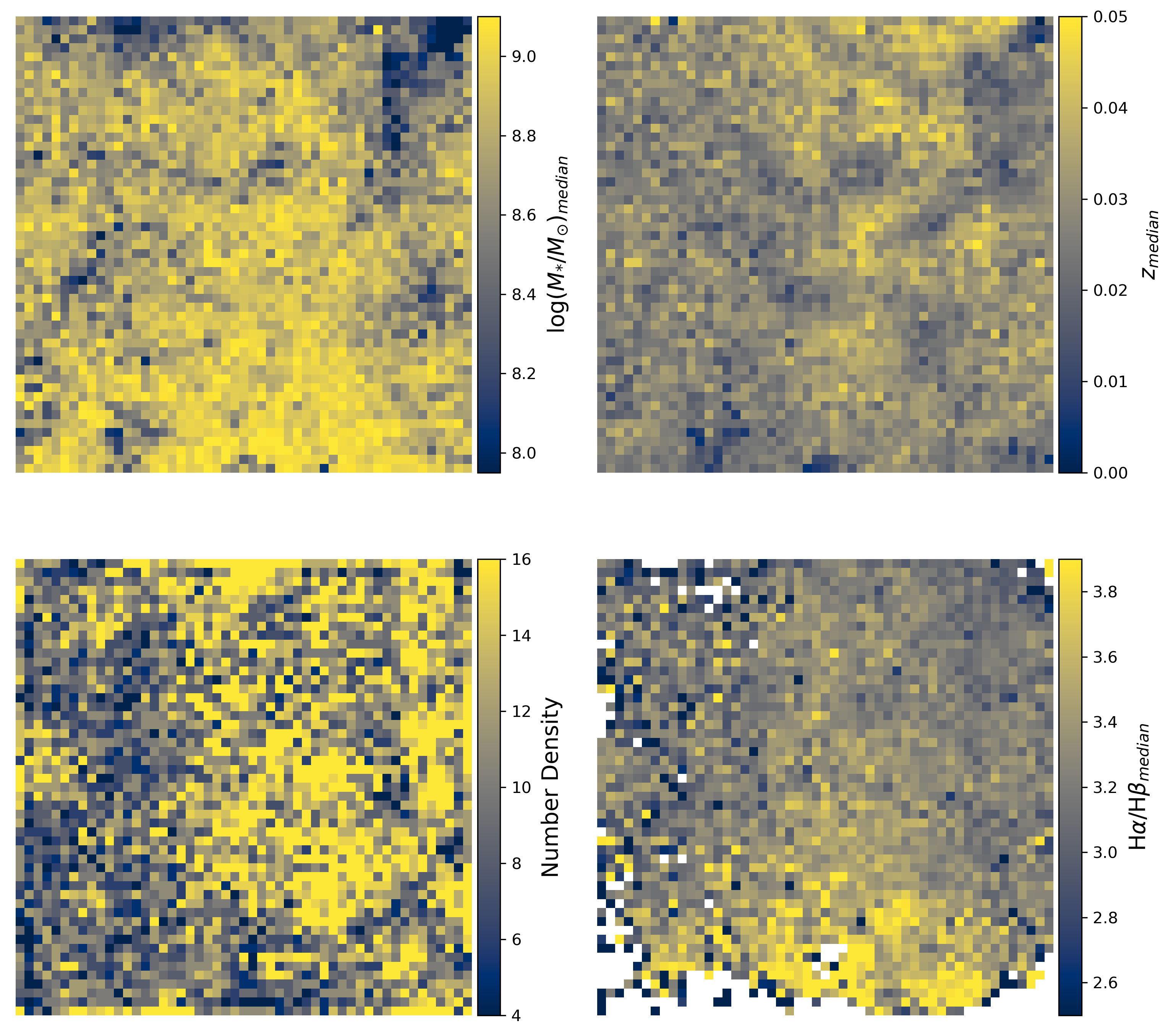}
\caption{SOM trained on the photometric colors of dwarf galaxies. 
\textbf{Top row:} SOM cells color-coded by the logarithmic median stellar mass (left) and photometric redshift (right) of the training galaxies mapped to each cell. 
\textbf{Bottom row:} SOM cells color-coded by the number density of training galaxies (left) and the median H$\alpha$/H$\beta$ line ratio (right). White regions along the lower edges of the SOM correspond to cells where H$\alpha$ or H$\beta$ measurements were unavailable.}

\label{fig:ND}
\end{figure*}

Figure~\ref{fig:ND} presents training-related results, highlighting the distribution of physical properties within the training sample of dwarf galaxies. The top-left panel shows the SOM color-coded by the median stellar mass of the training galaxies in each cell. As illustrated, the lowest-mass galaxies tend to cluster in specific regions on the upper left and right of the map, while the most massive systems are primarily located toward the bottom, reflecting a physically meaningful mass gradient across the SOM. The adjacent panel displays the distribution of photometric redshifts for the same training sample, allowing direct comparison with the stellar mass structure. Both stellar mass and redshift values are derived from the galaxies' SEDs, based on multi-band photometric fitting.
Another panel shows the number density of training galaxies mapped to each cell. This map confirms that nearly all cells are populated, suggesting that the predefined SOM size was well-chosen, large enough to minimize the grouping of dissimilar galaxies into the same cell, yet compact enough to avoid underutilized, empty regions of the map.

The final panel in Figure~\ref{fig:ND} presents the median H$\alpha$/H$\beta$ line ratio across the SOM, a proxy for dust attenuation. A clear gradient is evident, with higher values concentrated in the bottom regions of the map, coinciding with the most massive stellar populations, consistent with increased dust content in more evolved galaxies. The preservation of such topological relationships across multiple physical parameters demonstrates the effectiveness of the SOM in capturing structured variations in the input color space. The white regions along the lower edges of the SOM represent cells for which H$\alpha$ or H$\beta$ measurements were not available.



\section{Results and Discussion}

Dwarf galaxies in the NSA catalog serve as a valuable testbed for examining AGN selection across a broad range of galaxy properties, particularly at low stellar masses and luminosities. Unlike their more massive counterparts, dwarf galaxies typically host lower-mass black holes and exhibit higher levels of star formation and lower metallicities, all of which can obscure or mimic traditional AGN signatures. These conditions make AGN detection more challenging and selection methods more susceptible to contamination. By mapping previously identified AGN candidates onto the trained SOM, we investigate how different selection techniques, including optical emission line diagnostics, mid-infrared color cuts, broad-line features, variability, and X-ray emission, probe distinct regions of the photometric manifold in this low-mass regime. This approach allows us to assess both the complementarity and the limitations of existing AGN diagnostics in the dwarf galaxy population.
In this section, we use the reduced-dimensional projection of the learned SED space to visualize the selection functions of each AGN diagnostic across the NSA dwarf galaxy population. This framework facilitates a comprehensive comparison of AGN host properties and selection overlaps.

\subsection{Distribution of AGNs in Dwarf Galaxies}

\begin{figure*}[htbp]
\centering
  \includegraphics[trim=0cm 0cm 0cm 0cm, clip,width=1 \textwidth] {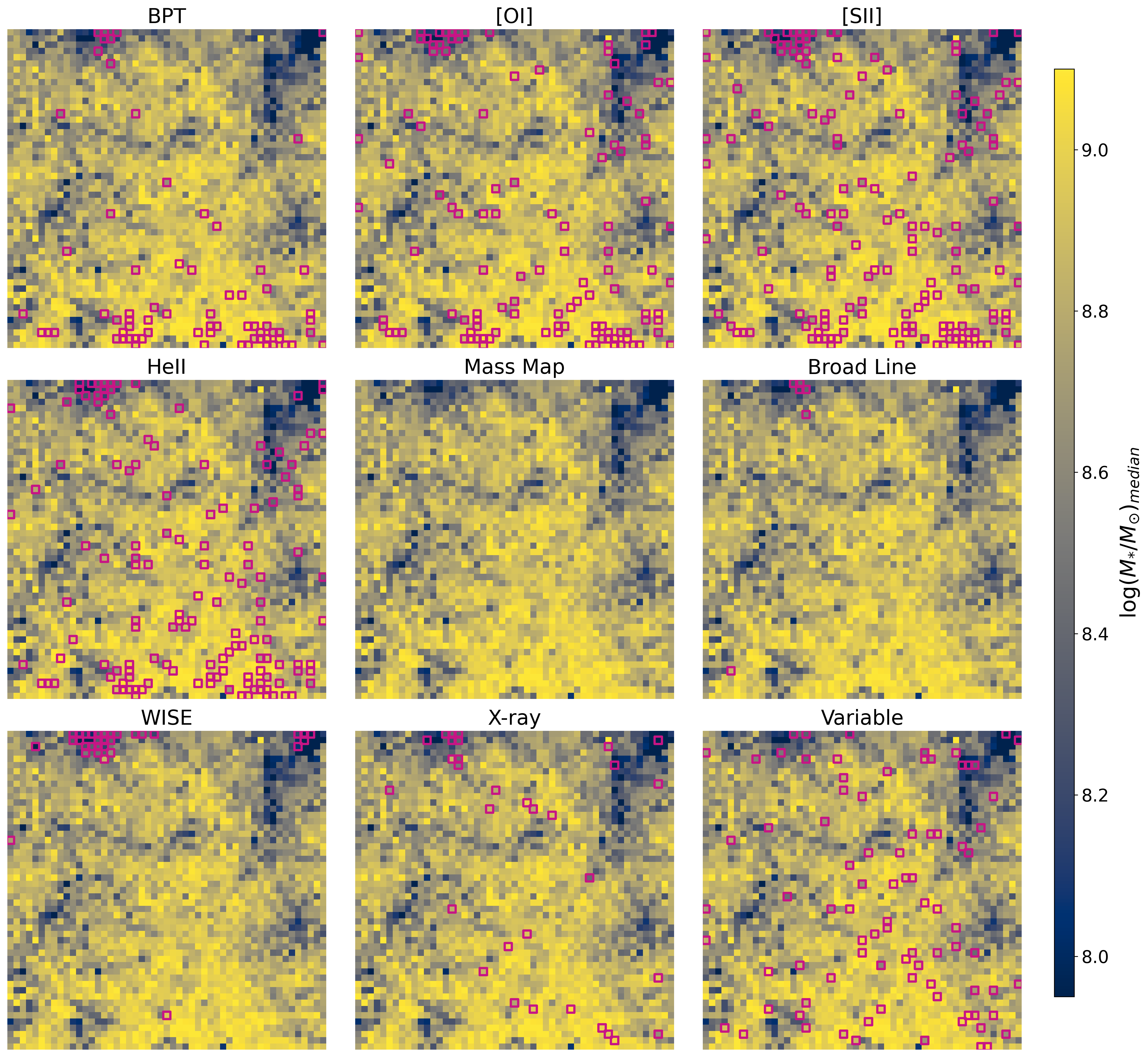}
\caption{Distribution of NSA dwarf AGNs mapped onto the trained SOM, with cells color-coded by the logarithmic median stellar mass of the training dwarf galaxies. 
Pink squares indicate AGNs associated with representative selection methods.
}

\label{fig:AGN}
\end{figure*}

\begin{figure*}[htbp]
\centering
  \includegraphics[trim=0cm 0cm 0cm 0cm, clip,width=1 \textwidth] {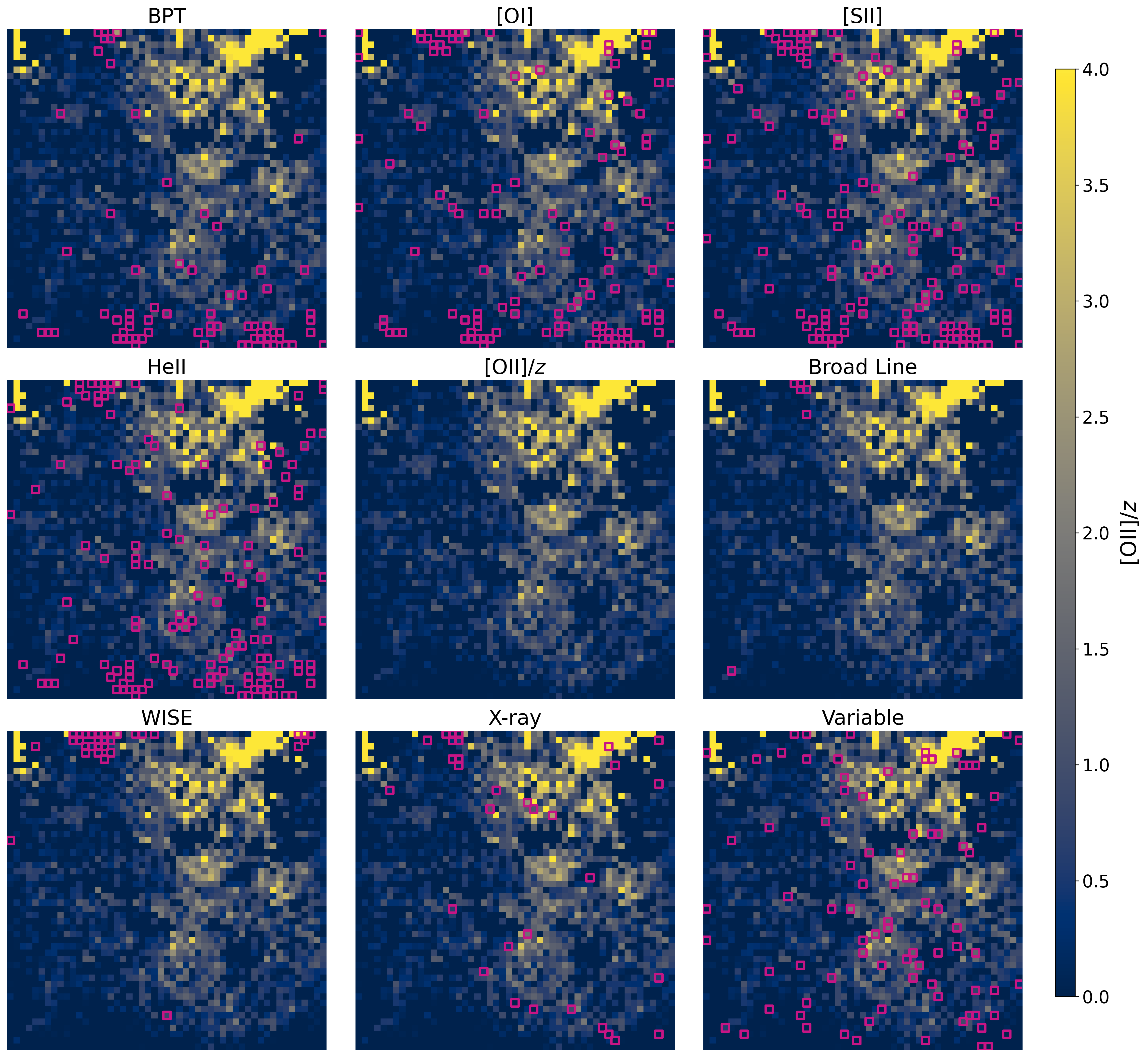}
\caption{Distribution of NSA dwarf AGNs mapped onto the trained SOM, with cells color-coded by the [O\,\textsc{ii}]/$z$ flux ratio of the training dwarf galaxies. The $z$ corresponds to SDSS $z$-band photometry, tracing the stellar continuum near 9000~\AA.
Pink squares indicate AGNs associated with representative selection methods.
}

\label{fig:o2z}
\end{figure*}

\begin{figure*}[htbp]
\centering
  \includegraphics[trim=0cm 0cm 0cm 0cm, clip,width=1 \textwidth] {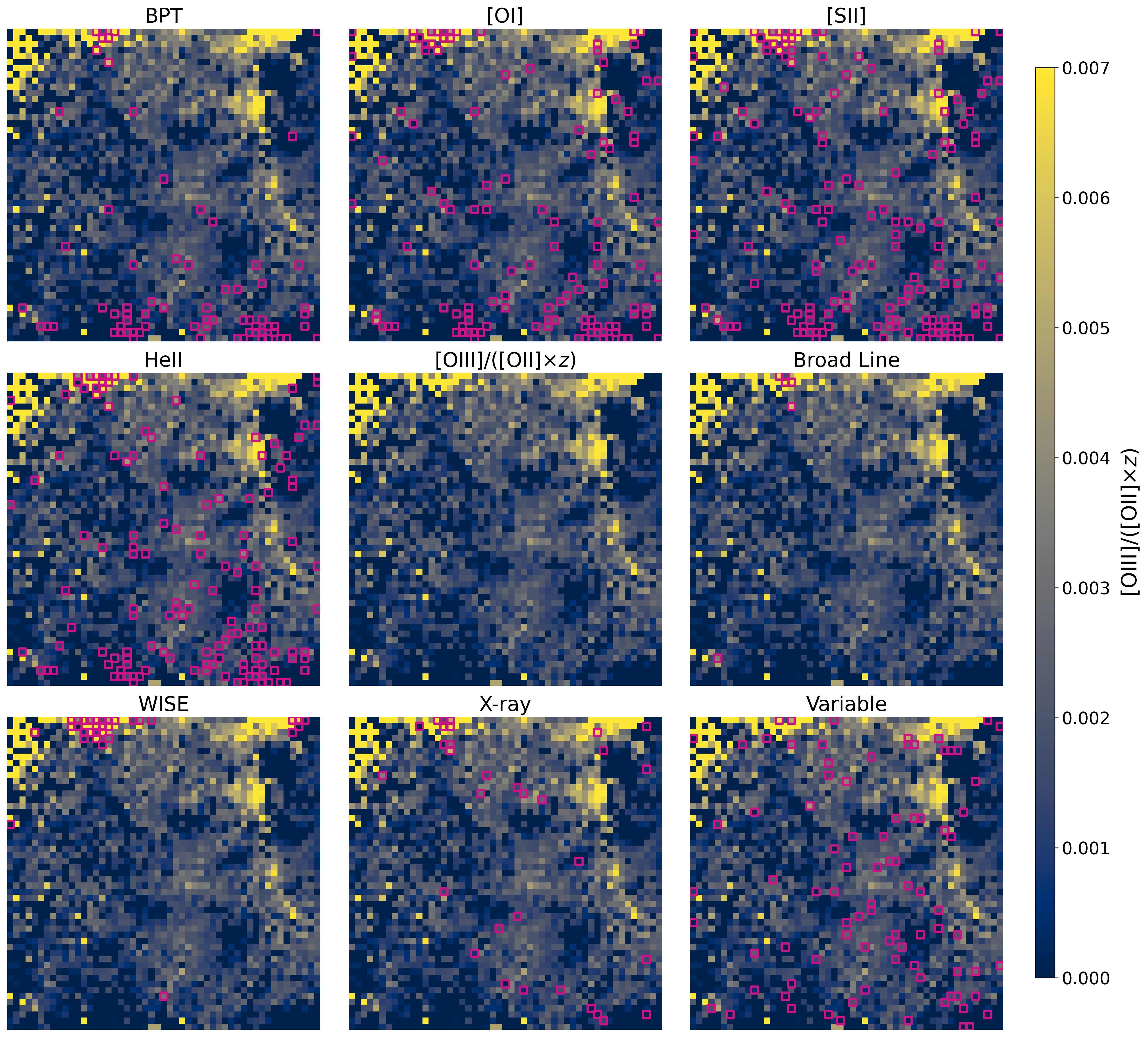}
\caption{Distribution of NSA dwarf AGNs mapped onto the trained SOM, with cells color-coded by the [O\,\textsc{iii}]/([O\,\textsc{ii}]~$\times$~$z$) flux ratio of the training dwarf galaxies. The $z$ corresponds to SDSS $z$-band photometry, tracing the stellar continuum near 9000~\AA.
Pink squares indicate AGNs associated with representative selection methods.
}

\label{fig:o32z}
\end{figure*}

The demographics of AGNs in dwarf galaxies remain poorly constrained, yet are essential for understanding the early formation and growth of black holes. Unlike their massive counterparts, dwarf galaxies are more likely to host low-mass black holes that have undergone minimal evolutionary processing, making them ideal laboratories for testing black hole seeding models. However, detecting AGNs in this regime is particularly challenging due to several physical factors, including their intrinsic faintness, elevated star formation activity, and low gas-phase metallicities. Low metallicity impacts AGN selection by altering the emission-line ratios commonly used in optical diagnostic diagrams such as the BPT \citep{Baldwin1981}, often pushing AGNs into regions typically associated with star-forming galaxies \citep{Groves2006, Feltre2016, Hirschmann2017, Cann2019}. This effect arises because lower metallicity gas has fewer coolants, resulting in harder ionizing spectra and stronger high-excitation lines even in purely star-forming regions, thereby increasing the likelihood of misclassification. Additionally, mid-infrared AGN selection can suffer from contamination by dusty starbursts, which are more common in metal-poor environments \citep{Satyapal2014}. As a result, current AGN samples in dwarf galaxies are incomplete and shaped by significant selection biases, limiting efforts to robustly determine black hole occupation fractions or scaling relations in the low-mass regime.

These challenges motivate the need for a unified framework to compare AGNs selected through different diagnostics. By projecting diverse AGN populations onto the same photometric SED space using a trained SOM, we can examine how each method samples the dwarf galaxy population and assess selection biases tied to host properties such as stellar mass, color, and redshift.

\begin{figure}[htbp]
\centering
  \includegraphics[trim=0cm 0cm 0cm 0cm,clip,width=0.5\textwidth] {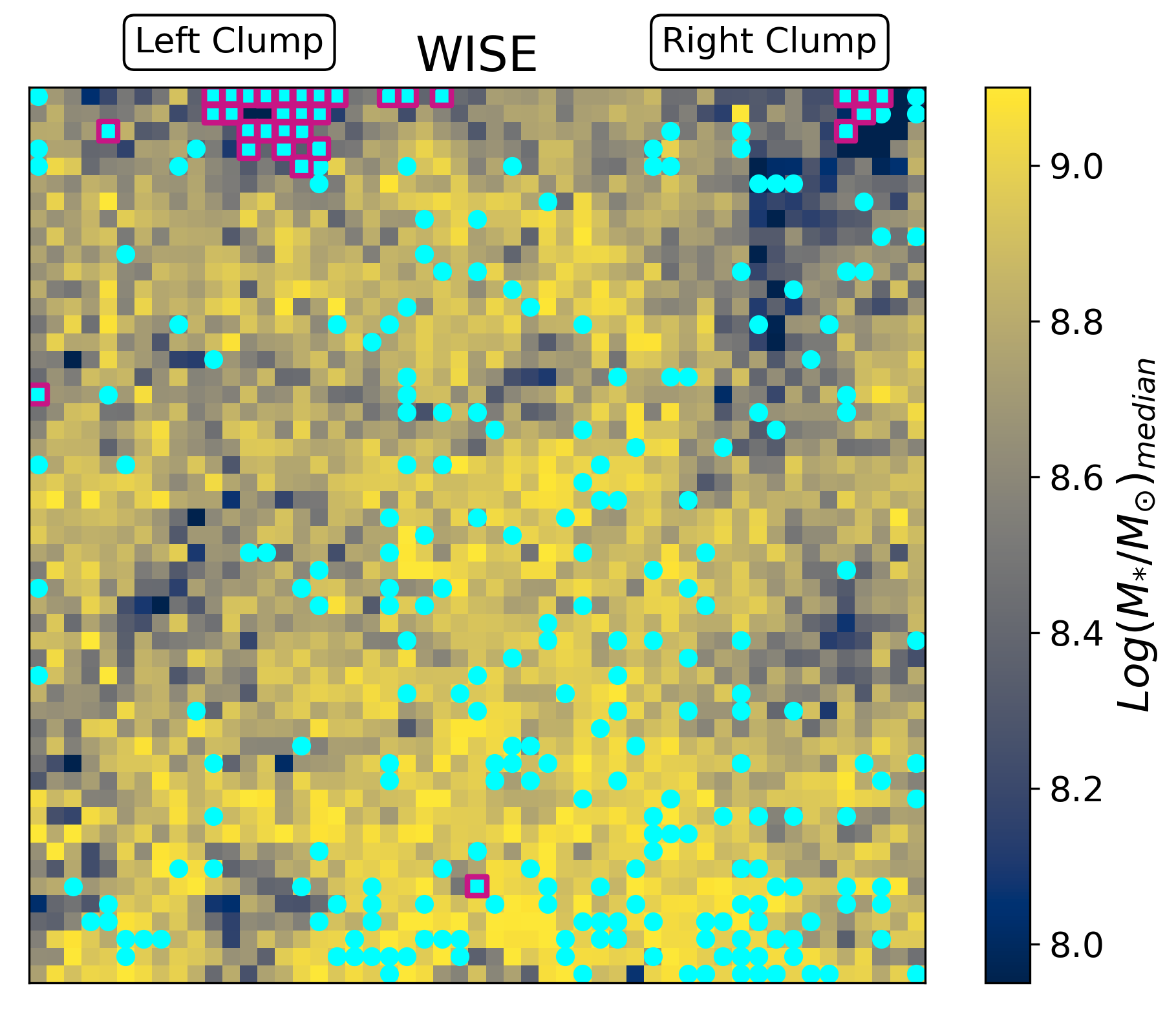}
\caption{WISE-selected dwarf AGNs mapped onto the trained SOM, shown as pink squares overlaid on cyan circles, which represent the full sample of NSA dwarf AGNs selected by various methods. Two distinct WISE AGN clumps are visible: the right clump is associated with bluer sources, while the left clump corresponds to redder AGNs. Among the 83 WISE-selected AGNs, 22 are mapped to the right clump and 55 to the left clump. The SOM background is color-coded by the logarithmic median stellar mass of the training galaxies in each cell.}

\label{fig:WISE}
\end{figure}

Figure~\ref{fig:AGN} shows AGNs selected by different diagnostic methods, each mapped separately onto the trained SOM. Pink squares highlight objects belonging to representative AGN selection groups, as indicated by the label on top of each panel. We repeat the mass map from Fig.~2 in the center of Fig.~\ref{fig:AGN} for comparison purposes. BPT-selected AGNs, based on traditional optical 
emission-line diagnostics using [O~\textsc{iii}]/H$\beta$ and [N~\textsc{ii}]/H$\alpha$ ratios \citep{Baldwin1981, Kewley2006}, appear clustered predominantly in the most massive regions of the SOM. This trend is consistent with known limitations of the BPT diagram in the low-mass regime, where a combination of weak emission lines, intense star formation, and low gas-phase metallicities reduces its effectiveness in distinguishing AGNs from star-forming galaxies \citep{Reines2013, Cann2019}. In particular, the [N~\textsc{ii}]/H$\alpha$ ratio becomes less reliable at low metallicities, often shifting AGNs into the star-forming region of the diagram, while strong nebular excitation from young, massive stars can mimic AGN-like [O~\textsc{iii}]/H$\beta$ values. As a result, the BPT diagram tends to miss AGNs in metal-poor, star-forming dwarfs and is biased toward more massive, higher-metallicity systems.

AGNs selected using the [O~\textsc{i}], [S~\textsc{ii}], and He~\textsc{ii} emission-line diagnostics exhibit an approximately uniform distribution across the SOM but show weak clustering trends toward less massive regions at the top and more massive regions at the bottom. In contrast, AGNs identified via X-ray and variability criteria show no apparent clustering, suggesting a truly uniform distribution. This indicates that these selection methods probe a wider range of SED shapes and are less strongly influenced by the host galaxy's mass. The widespread distribution of variability-selected AGNs across the manifold highlights the utility of this selection method for identifying AGNs across a broader range of galaxy properties. This has important implications for future time-domain surveys such as the Vera C. Rubin Observatory's Legacy Survey of Space and Time (LSST), which will be capable of detecting large, diverse populations of variable AGNs, including those in low-mass, obscured, or otherwise atypical hosts.

Notably, WISE-selected and broad-line AGNs are more concentrated in the lower-mass regions of the SOM, consistent with the notion that mid-infrared selection and spectroscopic broad-line identification are more effective at detecting AGNs in less massive galaxies \citep{Satyapal2014, Hainline2016, ReinesVolonteri2015}. However, the small number of broad-line AGNs in our sample highlights the limitations of this method, which is inherently incomplete due to its reliance on unobscured, high-luminosity nuclei with detectable broad emission lines \citep{Trump2011}.

Figure~\ref{fig:o2z} shows the distribution of NSA dwarf AGNs mapped onto the trained SOM, with cells color-coded by the [O\,\textsc{ii}]/$z$-band flux ratio of the training dwarf galaxies. The [O\,\textsc{ii}]~$\lambda3727$ emission line is a common tracer of star formation in galaxies (e.g., \citealt{Kennicutt1998}), while the SDSS $z$-band probes the stellar continuum near $\sim$9000~\AA, largely free of strong emission lines. This ratio serves as a proxy for specific star formation activity relative to stellar brightness.
Brighter regions in the SOM correspond to galaxies with stronger [O\,\textsc{ii}] emission relative to their stellar continuum — i.e., more starburst-like systems. As expected, AGNs selected via traditional diagnostics tend to avoid these regions of high [O\,\textsc{ii}]/$z$, reflecting the well-known difficulty in identifying AGNs in galaxies with intense star formation (e.g., \citealt{Kewley2001, Kauffmann2003}). In contrast, X-ray and optically variable AGNs are more frequently found in star-forming regions of the SOM. This is because their selection methods do not rely on optical emission-line diagnostics, which can be diluted or overwhelmed by strong nebular emission from H\,\textsc{ii} regions in high-SFR galaxies. X-ray emission traces high-energy accretion processes that are largely unaffected by host galaxy star formation or dust obscuration, while optical variability reflects intrinsic changes in the AGN accretion disk on short timescales and is independent of host galaxy line ratios. As a result, these AGNs are less biased against being identified in galaxies with strong ongoing star formation \citep[e.g.,][]{Hickox2009,Trump2015,SanchezSaez2018}.

Additionally, a comparison between Figure~\ref{fig:AGN} and Figure~\ref{fig:o2z} reveals that the most massive dwarf galaxies, clustered near the bottom of the SOM, exhibit the lowest [O\,\textsc{ii}]/$z$ values, consistent with lower star formation activity.

Figure~\ref{fig:o32z} presents the AGNs selected by various methods, overlaid on the SOM color-coded by the flux ratio [O\,\textsc{iii}]/([O\,\textsc{ii}]~$\times$~$z$). The [O\,\textsc{iii}]~$\lambda5007$ line is commonly used as a tracer of AGN activity (e.g., \citealt{Heckman2005}), while [O\,\textsc{ii}] traces star formation and $z$-band reflects stellar line-free continuum. Dividing [O\,\textsc{iii}] by $z$ provides an AGN luminosity measure relative to stellar light, and further dividing by [O\,\textsc{ii}] helps to mitigate contamination from star formation in the [O\,\textsc{iii}] flux. 

As illustrated in Figure~\ref{fig:o32z}, the brightest regions of the SOM correspond to galaxies hosting more luminous AGNs relative to the stellar mass of their host. 
Among the AGNs that cluster in the upper regions of the SOM—corresponding to lower-mass galaxies—a significant fraction fall in areas with elevated [O\,\textsc{iii}]/([O\,\textsc{ii}]~$\times$~$z$) values. This supports the interpretation that these methods (e.g., BPT, [O\,\textsc{i}], [S\,\textsc{ii}], He\,\textsc{ii}, broad-line, and WISE) tend to identify objects with higher ratios of AGN to stellar luminosities.

\begin{figure}[!htbp]
  \includegraphics[trim=0cm 0cm 0cm 0cm,clip,width=0.5\textwidth] {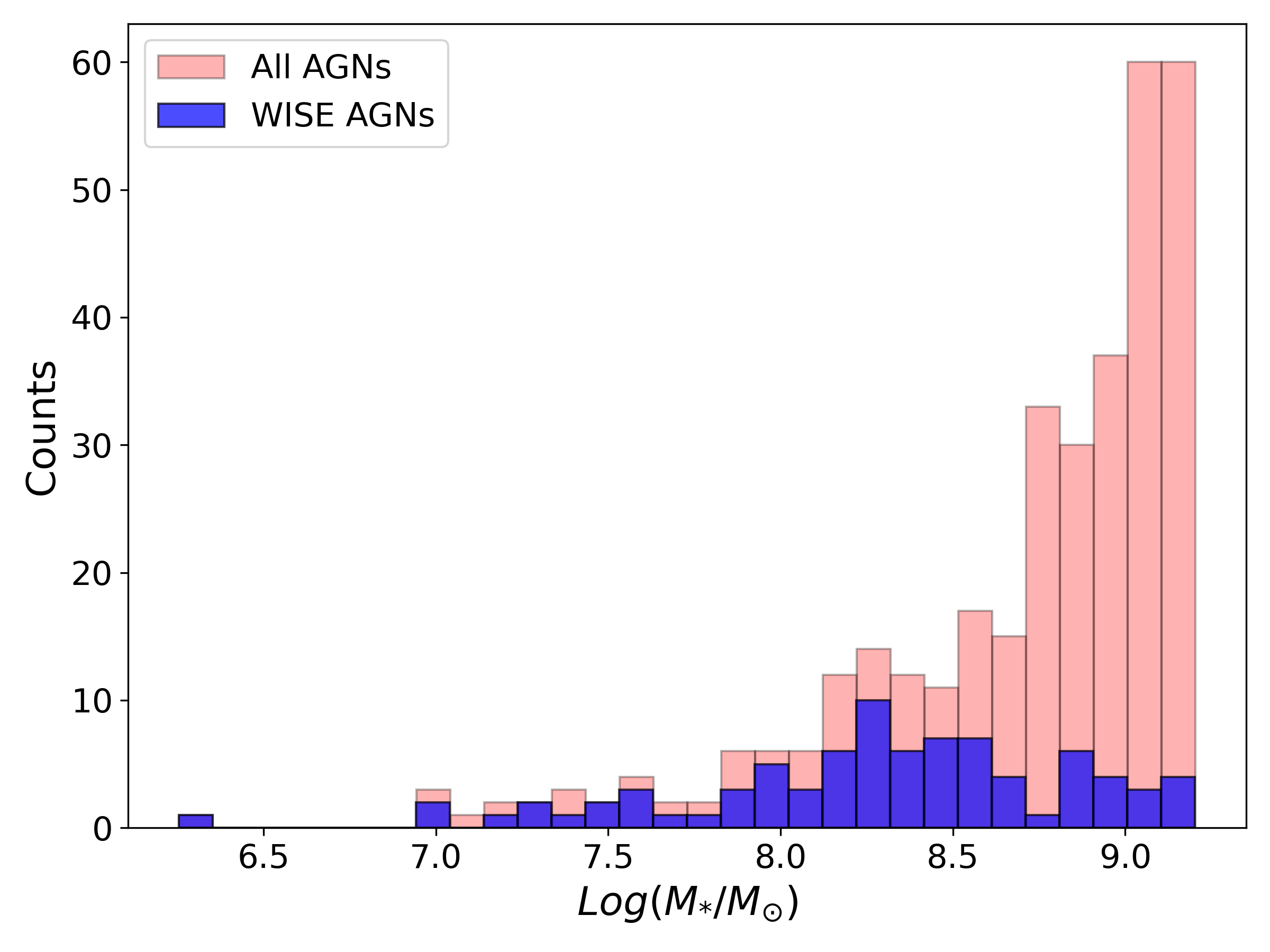}
\caption{Comparison of the stellar mass distributions of WISE-selected dwarf AGNs and those selected through other methods. WISE AGNs show a clear preference for lower-mass hosts relative to the broader dwarf AGN sample.}

\label{fig:mass}
\end{figure}

\begin{figure*}[!htbp]
   \includegraphics[trim=5cm 0cm 4.5cm 0cm,clip,width=1\textwidth] {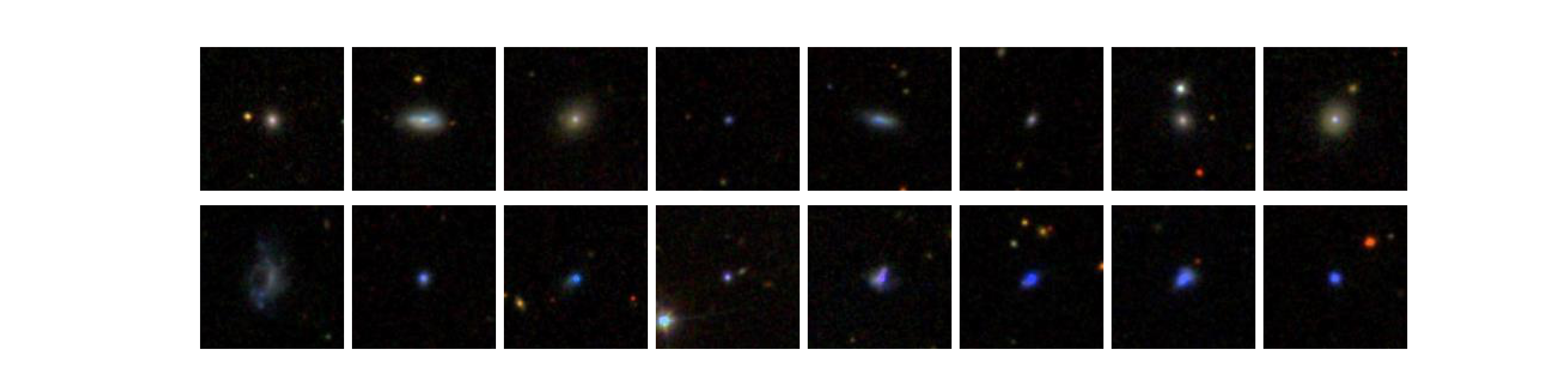}
 \caption{SDSS images of a sample of WISE-selected AGNs from the left (top) and right (bottom) clumps. The scale is 0.4"/pixel. Galaxies in the left clump appear less blue, and several of them have a bright source in the center of the galaxy. Galaxies in the right clump appear bluer, indicating a higher likelihood of starburst activity.}

 \label{fig:images}
 \end{figure*}

In general, as shown by the distribution of cyan circles in Figure~\ref{fig:WISE}, AGNs identified across all selection methods are dispersed throughout the SOM, with some clustering toward the bottom but overall widespread coverage. This spatial distribution reinforces the idea that AGN activity occurs across a broad diversity of host galaxies, spanning different stellar masses, colors, and histories, and is consistent with the picture in which AGNs flicker on and off in a large fraction of dwarf galaxies.

\subsection{WISE-selected Dwarf AGNs and Starburst Contamination}
Mid-infrared color selection using \textit{WISE} has become a widely used method to identify AGNs, particularly in systems where optical diagnostics may be obscured or unreliable. This approach relies on the redder mid-infrared colors produced by hot dust heated by an accreting black hole \citep{Stern2012, Satyapal2014}.

As shown in Figure~\ref{fig:WISE}, WISE-selected AGNs are strongly clustered in the lower-mass regions of the SOM, forming two distinct clumps on the top right and left sides of the map. This spatial concentration reflects the tendency of WISE selection to favor lower-mass galaxies, possibly due to enhanced dust-heated mid-infrared emission in compact or gas-rich systems \citep{Hainline2016}. The fact that these AGNs occupy the low-mass end of the SED manifold is consistent with their stellar mass distribution (Figure~\ref{fig:mass}), which skews lower compared to AGNs selected through other methods. This bias is further supported by studies showing that infrared-based selection techniques are more effective at identifying high-luminosity AGNs in low-mass galaxies, where the AGN can outshine the host in the IR \citep{Messias2013}.
The location of the WISE clumps (Figure~\ref{fig:WISE}) corresponds well to the regions of the SOM color-coded by W1$-$W2, which exhibit higher values of this mid-infrared color, in accordance with expectations, since W1$-$W2 was used as part of the input for training (Figure~\ref{fig:som_train}). As a result, the SOM mapping process is naturally sensitive to variations in this color, enhancing the separation of WISE-selected sources. At first glance, the observed clustering may appear to be fully driven by the inclusion of WISE bands. However, after retraining the SOM without the WISE photometry, we found that the WISE-selected sources still occupied broadly similar regions of the manifold, although the distinct clumpy structure became less concentrated. This indicates that the grouped sources also share coherent near-UV to optical SED features, and that the WISE bands further sharpen the distinction by highlighting dust-heated mid-infrared emission.

While WISE color selection has proven effective at identifying luminous AGNs \citep{Jarrett2013}, it suffers from significant contamination by dusty starburst galaxies in the dwarf galaxy regime \citep{Hainline2016}. This is primarily because intense star formation in low-mass galaxies can heat dust and produce mid-infrared colors similar to those of AGNs, particularly in the W1 (3.4~µm) and W2 (4.6~µm) bands used for AGN selection. Moreover, the lower metallicities typical of dwarf galaxies can alter the dust-to-gas ratio and the shape of the infrared SED, enhancing this degeneracy \citep{Hainline2016, Satyapal2014}. As a result, WISE color cuts, such as those based on W1$-$W2, may misclassify compact, star-forming systems as AGN hosts. 

\begin{figure*}[!htbp]
  \includegraphics[trim=0.4cm 0cm 0cm 0.2cm,clip,width=0.5\textwidth] {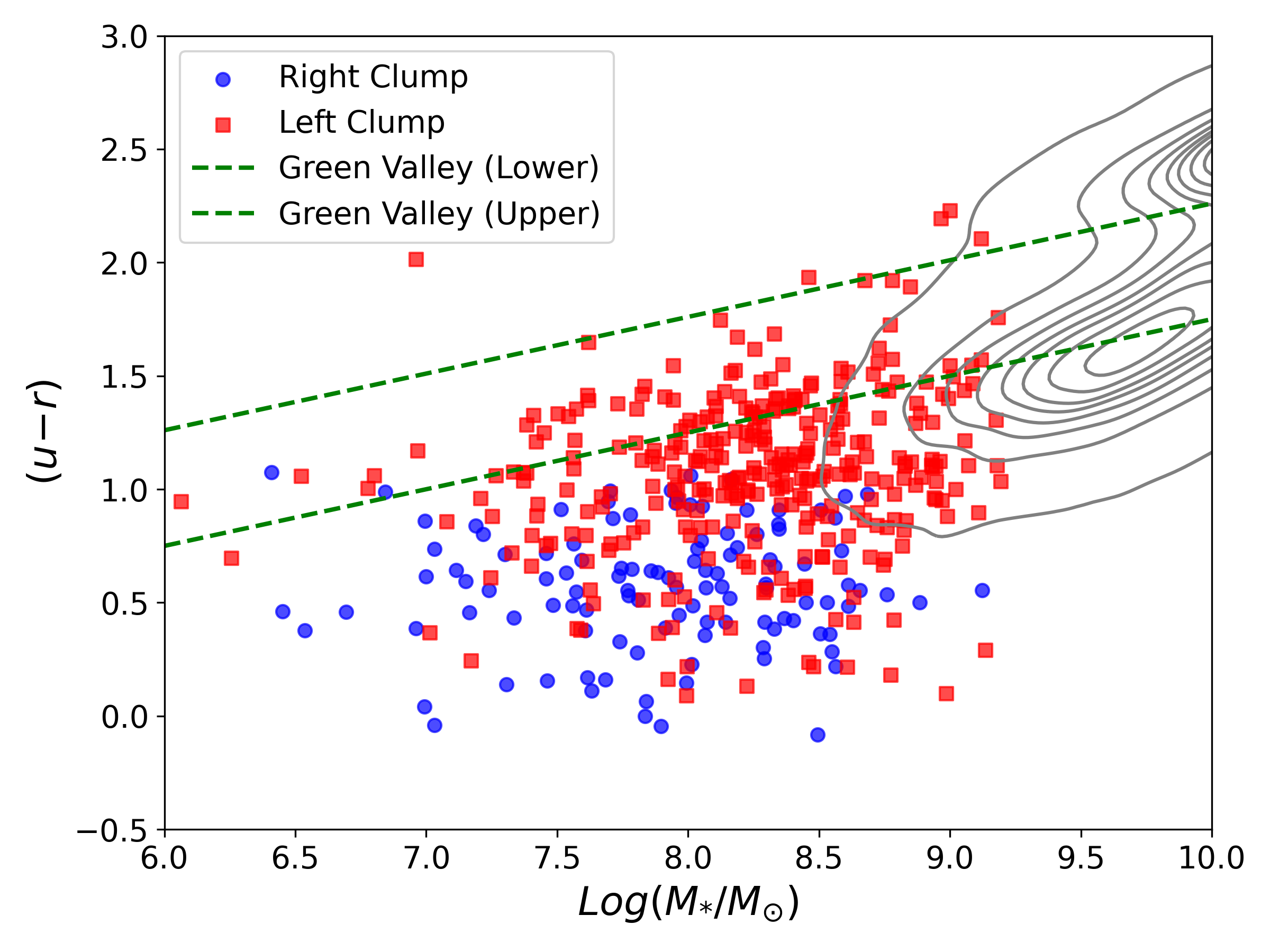}
  \includegraphics[trim=0.4cm 0cm 0cm 0.2cm,clip,width=0.5\textwidth] {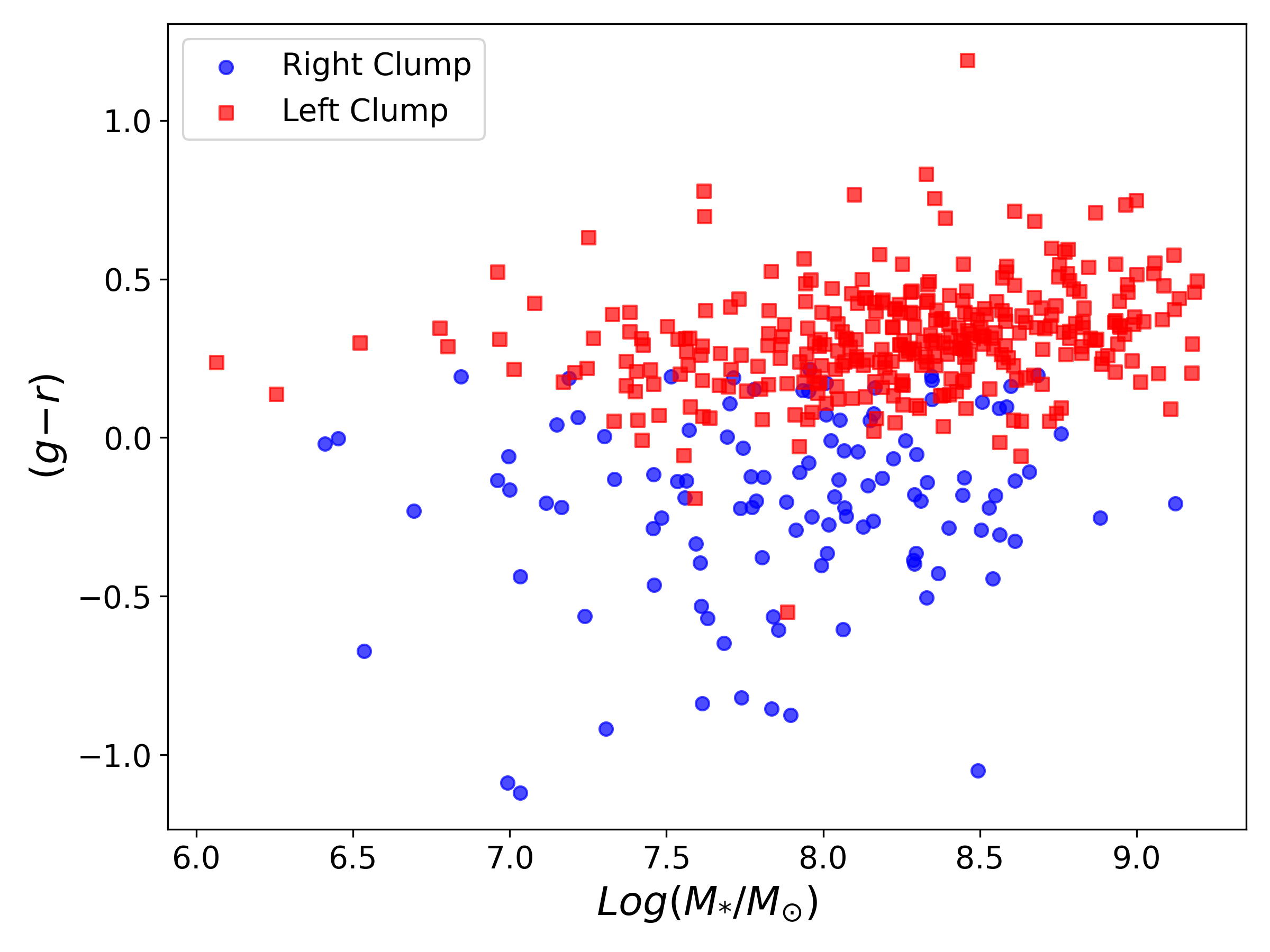}
\caption{Color–mass diagrams of all training and test sample galaxies located within the left and right WISE clump cells of the SOM. The bluer $u\!-\!r$ colors of the right clump galaxies suggest a population dominated by young starbursts, consistent with contamination in WISE-based AGN selection. The contours and the upper and lower boundaries of the green valley shown in the left panel are adopted from \citet{Schawinski2014}, who define this transition region in the $u\!-\!r$ versus stellar mass plane based on a volume-limited sample of galaxies from SDSS. These reference lines help distinguish between blue cloud, green valley, and red sequence populations.
}

\label{fig:colormass}
\end{figure*}

In our analysis, the presence of two separate WISE-selected AGN clumps in the SOM offers a potential way to disentangle true AGN hosts from starburst contaminants. These clumps are located in the upper right and upper left of center  parts of the SOM, respectively, as labeled on Fig.~\ref{fig:WISE}, and hereafter referred to as ``right'' and ``left'' clumps. Notably, the left clump exhibits a stronger overlap with AGNs identified by multiple independent diagnostics, including BPT, broad-line, X-ray, and narrow emission-line selections such as [O\,\textsc{i}], [S\,\textsc{ii}], and He\,\textsc{ii}, compared to the right clump (Figures~\ref{fig:AGN}, \ref{fig:o2z} and \ref{fig:o32z}). Among these, WISE-selected AGNs show the most prominent presence in the left clump, with 55 WISE AGNs located in this region. For comparison, the [S\,\textsc{ii}]-based selection, which shows the largest overlap with WISE in this clump, identifies 22 AGNs in the same area. This convergence of multiple AGN indicators within a specific region of the SOM suggests a higher likelihood of true AGN activity in that area, reflecting consistent underlying physical properties across selection methods.

Moreover, the left and right clumps differ in terms of their associated line-ratio flux distributions. As shown in Figure~\ref{fig:o2z}, several WISE-selected AGNs in the right clump are located in regions of the SOM with elevated [O\,\textsc{ii}]/$z$ flux ratios, indicative of enhanced star formation or starburst-like activity. In contrast, AGNs in the left clump tend to avoid these high [O\,\textsc{ii}]/$z$ regions, suggesting lower levels of star formation in their host galaxies.

Figure~\ref{fig:o32z} further highlights this distinction: most AGNs in the left clump are concentrated in SOM regions with the highest [O\,\textsc{iii}]/([O\,\textsc{ii}]~$\times$~$z$) flux ratios, consistent with powerful AGNs in relatively faint, low-mass galaxies. In contrast, the right clump contains fewer AGNs in these high-ratio regions, and some of its cells tend to avoid them entirely.
Furthermore, Visual inspection of SDSS imaging (Figure~\ref{fig:images}) for galaxies in each clump reveals that the right clump predominantly consists of bluer WISE-selected sources, consistent with ongoing star formation, while the left clump is comprised of redder sources that largely lack visible signs of recent star-forming activity. This pattern suggests that the right clump of WISE-selected AGNs may include a higher fraction of star-forming contaminants, possibly due to mid-infrared emission from intense star formation mimicking AGN-like colors.

Further support comes from mapping all SOM training galaxies in the same cells as the WISE-selected AGNs onto color–mass diagrams (e.g., $(u - r)$ and $(g - r)$ versus stellar mass; Figure~\ref{fig:colormass}). The galaxies associated with the right clump tend to fall in regions characteristic of intense star formation, whereas those in the left clump occupy redder areas. These results suggest that the SOM may naturally separate WISE-selected AGNs into two populations: those likely dominated by starburst emission, and those more likely to host accreting black holes.

\subsection{Coronal Line Emitters}

Coronal emission lines, which arise from highly ionized species such as [Ne~\textsc{v}], [Fe~\textsc{x}], and [Si~\textsc{vi}], are among the most compelling indicators of active galactic nuclei (AGN) activity. These lines are often referred to as the ``smoking gun'' of accreting black holes due to their requirement for an intense, hard ionizing radiation field \citep{Rodriguez2002, Riffel2006, Rodriguez2011, Mueller2011, Lamperti2017}. Their detection in dwarf galaxies provides particularly valuable evidence for the presence of low-mass black holes, offering a unique window into black hole growth at the low-mass end of the galaxy population 
\citep{Cann2019, Aravindan2024, Reefe2022}.

To investigate this population in our sample, we cross-matched the full NSA dwarf galaxy catalog with previously published lists of sources exhibiting optical or near-infrared coronal line emission in dwarf galaxies  \citep{Bohn2021, Molina2021, Reefe2022}. This search yielded 57 dwarf galaxies with detected coronal lines. Of these, 48 were uniquely identified and notably absent from the other 
AGN classifications in \citet{WasleskeBaldassare2024}. 

\begin{figure}[!htbp]
\centering
  \includegraphics[trim=0cm 0cm 0cm 0cm,clip,width=0.5\textwidth] {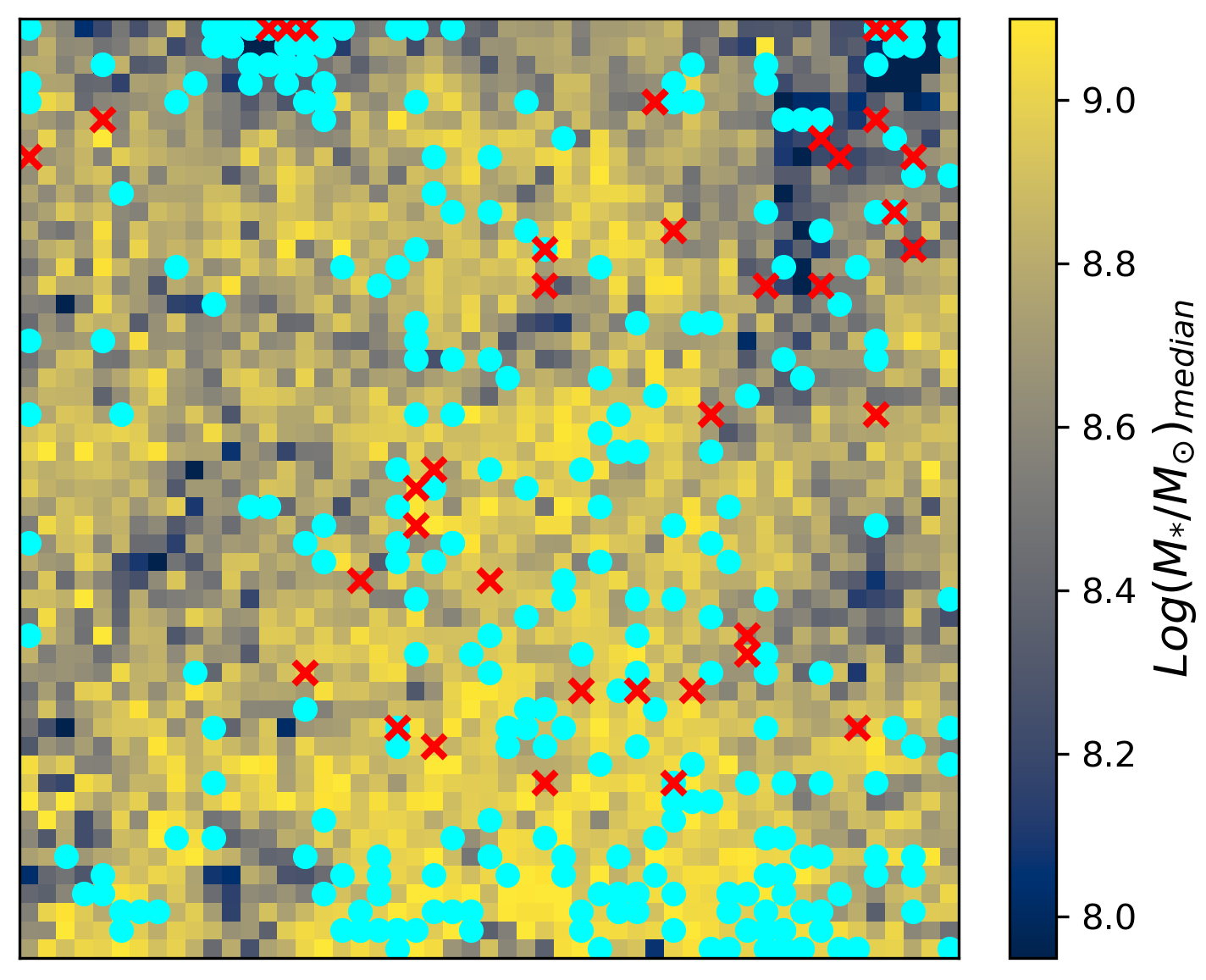}
\caption{Distribution of NSA dwarf AGNs and coronal line emitters mapped onto the trained SOM, which is color-coded by the logarithmic median stellar mass of the training galaxies in each cell. Cyan circles indicate the mapped AGNs, while red crosses show the coronal line emitters.}

\label{fig:coronal}
\end{figure}

We mapped these coronal-line emitters onto our trained SOM, as shown in Figure~\ref{fig:coronal}, where they are denoted by red X symbols. While the clustering is not particularly strong, coronal-line emitters in our sample tend to avoid the bottom regions of the SOM, associated with the most massive galaxies, and are more frequently mapped to cells corresponding to lower-mass systems. The median logarithmic stellar mass of the SOM cells hosting these sources is 8.69, compared to 9.00 for the most massive regions of the map, supporting a mild preference for lower-mass environments. This trend reflects a number of physical factors: the reduced gravitational radius of low-mass black holes brings the coronal-line emitting region closer to the accretion engine, increasing the ionization parameter and enhancing the visibility of these high-excitation features. Compact nuclear environments in dwarf galaxies can further sustain extreme ionization conditions, particularly during episodes of radiatively efficient accretion that may approach high Eddington ratios, even at relatively modest bolometric luminosities \citep{RodriguezArdila2002, Riffel2006, RodriguezArdila2011, Lamperti2017, Aravindan2024}. Moreover, the reduced levels of gas and dust in such systems minimize nuclear obscuration, and a weaker stellar continuum lessens dilution effects in integrated spectra, improving detectability \citep{Cann2019}.

While the observed distribution reinforces the idea that coronal-line AGNs are prevalent in low-mass galaxies and potentially linked to unique accretion or ionization conditions in this regime, we caution that our sample is incomplete. The current census of coronal-line emitters is biased toward previously reported cases, which depend on the availability and sensitivity of existing optical and near-infrared spectroscopic data. It is likely that additional coronal-line AGNs remain undetected due to observational limitations. Expanding this census will require deeper, targeted spectroscopy, particularly in low-luminosity and obscured systems where coronal features may be faint or partially blended. Future surveys with enhanced sensitivity to high-ionization lines could substantially improve our understanding of this rare but important population.



\section{Summary}

In this study, we applied unsupervised machine learning to investigate the demographics and selection biases of AGNs in dwarf galaxies. We used SOM to analyze the high-dimensional SED space of dwarf galaxies, leveraging their ability to preserve topological structure in a lower-dimensional grid \citep{Kohonen1982, Geach2012, Hemmati2019, Sanjaripour2024}. Our SOM was trained using 30,344 dwarf galaxies from the NSA catalog version \texttt{nsa\_v1\_0\_1} (\( z < 0.055 \), \( M_* < 10^{9.5} M_{\odot} \)) in a 7-dimensional color space derived from 9-band photometry ranging from near-UV to mid-infrared. We then mapped 438 dwarf galaxies known to host  AGNs - selected through a variety of methods compiled by \citet{Wasleske2024}  to the trained SOM to study their distribution and underlying host galaxy properties.

\vspace{0.5em}
\noindent
Our key findings are summarized below:
\begin{itemize}
    \item The trained SOM preserves physical correlations in the data, such as the relationship between SED shape, stellar mass, and H$\alpha$/H$\beta$ line ratio (Figures~\ref{fig:som_train} and~\ref{fig:ND}).

    \item AGNs selected through different methods populate distinct and sometimes overlapping regions of the SOM, highlighting the biases and incompleteness of individual selection techniques (Figure~\ref{fig:AGN}).

    \item BPT-selected AGNs are confined to more massive regions of the SOM, reflecting their inefficiency in identifying low-mass AGNs due to weak or absent emission lines in metal-poor hosts (Figure~\ref{fig:AGN}).

    \item AGNs selected via [O~\textsc{i}], [S~\textsc{ii}], and He~\textsc{ii} diagnostics exhibit broadly distributed positions on the SOM, with weak clustering toward lower-mass regions at the top and higher-mass regions at the bottom. In contrast, X-ray and variability-selected AGNs show no apparent clustering, suggesting that these methods trace AGNs across a wider range of host galaxy properties (Figure~\ref{fig:AGN}).
    
    \item The uniform distribution of variability-selected AGNs across the SOM suggests that upcoming time-domain surveys, such as those from the Rubin Observatory (LSST), will be well-positioned to identify AGNs across a wide range of host galaxy properties, offering a largely unbiased census of the variable AGN population (Figure~\ref{fig:AGN}).

    \item AGNs selected via traditional emission-line diagnostics (e.g., BPT, [O\,\textsc{i}], [S\,\textsc{ii}], He\,\textsc{ii}), as well as broad-line and WISE methods, tend to avoid regions of the SOM with high [O\,\textsc{ii}]/$z$ flux ratios, consistent with the difficulty of identifying AGNs in strongly star-forming galaxies where nebular emission dominates (Figure~\ref{fig:o2z}).

    \item Among the AGNs that occupy the upper region of the SOM, corresponding to lower-mass galaxies, a significant fraction fall within the brightest areas of the SOM color-coded by [O\,\textsc{iii}]/([O\,\textsc{ii}]~$\times$~$z$), highlighting luminous AGNs residing in faint, low-mass hosts; these include sources selected via BPT, [O\,\textsc{i}], [S\,\textsc{ii}], He\,\textsc{ii}, broad-line, and WISE methods (Figure~\ref{fig:o32z}).

    \item WISE-selected AGNs are consistently mapped to the lower-mass regions of the SOM, confirming previous findings that mid-infrared color selection is effective for identifying AGNs in dwarf galaxies (Figures~\ref{fig:WISE} and~\ref{fig:mass}).
    
    \item These WISE-selected AGNs form two distinct clumps within the SOM: a right-side clump dominated by bluer, starburst-like galaxies, and a left-side clump composed of redder systems with spectral features more consistent with AGN activity. This separation highlights the potential of SOM to distinguish true AGN hosts from starburst contaminants within WISE-selected samples. (Figure~\ref{fig:WISE})

    \item Coronal-line emitters mapped onto the SOM generally avoid the most massive regions, suggesting a preference for lower-mass hosts. However, the current sample is incomplete and likely biased by limited spectroscopic coverage, underscoring the need for deeper, targeted observations to uncover the full population (Figure~\ref{fig:coronal}).

\end{itemize}

\section*{Acknowledgements}

We thank Michael McDonald and Nahum Arav for helpful discussions, and Chengzhang Jiang for sharing ideas relevant to this work. We also acknowledge our collaborators for their input in shaping the direction of the project.   Partial support for A.A.\ and G.C.\ was provided by NASA through a grant for program JWST-GO-03663 from the Space Telescope Science Institute, which is operated by the Association of Universities for Research in Astronomy, Inc., under NASA contract NAS 5-03127.

\bibliography{references.bib}

\end{document}